\documentclass[12pt]{article}
\usepackage{epsfig}

\parskip=\medskipamount 
plus.3em minus.5em \abovedisplayshortskip=.5em plus.2em minus.4em
\renewcommand{\thesection}{\arabic{section}}

\catcode`@=11

\@addtoreset{equation}{section} \@addtoreset{equation}{subsection}
\def\theequation{\ifnum\value{section}=0 \arabic{equation}\ignorespaces
\else \ifnum\value{section}=-1 A.\arabic{equation}\ignorespaces
\else \ifnum\value{subsection}=0
\thesection.\arabic{equation}\ignorespaces \else
\thesection.\arabic{subsection}.\arabic{equation}\ignorespaces
                             \fi
                        \fi
                   \fi}

{\catcode`\'=\active \def'{{}^\bgroup\prim@s}}

\catcode`@=12

\newcommand{\bq}{\begin{equation}}
\newcommand{\be}{\begin{equation}}
\newcommand{\fq}{\end{equation}}
\newcommand{\ee}{\end{equation}}
\newcommand{\bqr}{\begin{eqnarray}}
\newcommand{\beqs}{\begin{eqnarray}}
\newcommand{\fqr}{\end{eqnarray}}
\newcommand{\eeqs}{\end{eqnarray}}

\newcommand{\rf}[1]{(\ref{#1})}

\def\bop#1{\setbox0=\hbox{$#1M$}\mkern1.5mu
    \vbox{\hrule height0pt depth.04\ht0
    \hbox{\vrule width.04\ht0 height.9\ht0 \kern.9\ht0
    \vrule width.04\ht0}\hrule height.04\ht0}\mkern1.5mu}

\begin{document}
\thispagestyle{empty}

\begin{flushright}
\begin{tabular}{l}
hep-th/0506012 \\
\end{tabular}
\end{flushright}

\vskip .6in
\begin{center}

{\bf  Inversions of the Effective Action in Condensed Matter Models}

\vskip .6in

{\bf Gordon Chalmers}
\\[5mm]

{e-mail: gordon@quartz.shango.com}

\vskip .5in minus .2in

{\bf Abstract}

\end{center}
  
The quantum effective action may be used to invert information from 
phenomena, either measured or ideal, to the microscopic Lagrangian.  
As an example of this procedure the lattice composition of a solid 
can be determined in principle from desired critical phenomena.  
Also, the engineering of scattering and optical properties of 
particular degrees of freedom, and excitations, in terms of the 
temperature and electro-magnetic fields may be performed.  

\vfill\pagebreak

Quantum field theory in general can be 'solved' in terms of the 
functional integration of the quantum fields.  One recipe for doing 
this involves the integration of the propagating modes in perturbation 
theory; the perturbative aspect can be changed once the coupling 
dependence of series is obtained.  A procedure for obtaining the 
full scattering, and the quantum effective action that generates it, 
is described in \cite{ChalmersOne}-\cite{ChalmersThree} (and references 
therein); scalar, gauge, and gravitational models 
are analyzed within the perturbative context in \cite{ChalmersOne}, 
\cite{ChalmersTwo} and classically in \cite{ChalmersFive} and 
\cite{ChalmersSix}.  Further simplifications of the tensor algebra may 
be performed \cite{ChalmersFour}.

The quantum action of condensed matter models can also be obtained 
with the full dependence on the coupling constants, as a power series.  
The lattice type and composition is incorporated into the propagators 
and integration region (via the Brillioun zone, for example).   For 
example, the action describing phonons and electrons in the non-relativistic 
limit can be obtained to all orders in the coupling and energy scales. 

The quantum action as a function of the lattice $\Lambda$, the structure 
constants and properties of the cores $f_i$, non-linearities associated 
with the electrons and the band, the temperature $T$, and external fields 
$E$ and $B$ may be included into the determination of the quantum action.  
The microscopic coupling constants in the bare Lagrangian can be given 
their full perturbative series.   

The effective action in the condensed matter context takes on the form, 

\bqr 
\psi \Delta(x;T,\Lambda)\psi + m\psi\psi + F^2_{\rm phonon} + F^2_{\rm e.m.} 
\fqr 
together with the interactions, 

\bqr 
\prod \psi \psi \ldots 
  \psi A_{\rm p} \ldots A_{\rm p} A_{\rm e.m.} \ldots A_{\rm e.m.} 
\label{particlecombos}
\fqr 
combined with the derivatives placed in physically allowed places, which 
are derived from the fundamental theory.  The bare theory could also 
include the electromagnetic interactions derived from the atomic cores, 
and the non-linearities associated with the phonon displacements (i.e. 
quartic and higher order terms).  

The quantization of the bare theory should take into account the Brillouin 
zone features in the $k$-space integration.  Within the derivative expansion, 
this integration change does not alter the complications with the integrals.  

The application of the quantization, or mode expansion in the classical 
and quantum regime, generates the coefficients of the terms in 
\rf{particlecombos} in momentum space.  These coefficients, 

\bqr 
f_{s_{ij};f_i,\Lambda,T;g_j}
\fqr 
are expansions in the lattice structure $\Lambda$, $f_i$ and the coupling 
constants $g_j$.  The latter model the electron, phonon, and electromagnetic 
effects; they could also represent the core potentials of the atoms and 
non-linearities associated with the lattice.  These functions are computable 
within the derivative expansion to any order once the bare Lagrangian is 
specified.  (A momentum representation could be used to formulate the 
interaction potential from the cores in the lattice.  The temperature 
dependence is also included in the bare theory.)

The integral representations in \cite{ChalmersOne}-\cite{ChalmersTwo} can 
be used to find the 
full effective action and the coefficients $f_{s_{ij};f_i,\Lambda,T;g_j}$ 
as power series.  All couplings $f_i$, $\Lambda$, $g_i$ and the temperature 
$T$ can be left as arbitrary parameters in the bare theory.  The phonon 
(and other modes) populations at a given temperature can be inserted as 
background fields. 

The point of this calculation is that with the arbitrariness of these 
parameters, the functions $f$ multiplying the effective action's terms 
can be used to deduce the parameters.  In a quantum field theory form, 
or at a point at criticality in the lattice model, this inversion is 
clear from the power series form in the coupling as computed in 
\cite{ChalmersOne}-\cite{ChalmersSix}.  

It would be useful to compute the entire effective action as a function 
of all of these bare interactions, and to produce a computational program 
to carry out the inversion.  This would enable a reverse modeling, that 
is, to model desired phenomena even at non-criticality, to the bare 
Lagrangian including the construction of the lattice and its constituents.  
This program is technically straightforward to carry out in practice, 
without complications in principle.  The full effective action is available 
via the summation of Feynman diagrams, or those in the derivative expansion.  
The inversion of the functions $f_{s_{ij};\Lambda,T,g_i}$ to the couplings 
and lattice involve polynomial and transcendental equation solutions.  
However, in principle, any desired phenomena at or near criticality can 
be obtained with enough self-consistent information from the couplings and 
lattice.    

The band electron band information can be obtained from the low-energy 
effective action, including the quantum propagator.  Associated information 
from an external photon field can be used further to turn on background 
couplings and change further these functions $f$ by including the $\langle 
A_{\rm e.m.} \ldots A_{\rm e.m.}\rangle$.  The momentum dependence (or 
frequency bandwith and amplitude) of the external field $A_{\rm e.m.}$ can 
be used in theory to alter a near infinite number of terms, depending on the 
precision and accuracy of the waveform.  These feature of the backgound 
electromagnetic field can be used to alter substantially the band structure 
of the electrons and the Fermi filling (after including possible interactions 
with the lattice cores).  

The inversion of the effective action can be done with 
the background fields to obtain a variety of phenomena.  Even seemingly 
complicated lattices, such as the cuprates, can be modeled in a 
straightforward fashion, once the bare Lagrangian is produced.  Bandwidth 
aborption and reflection can be produced as a function of desired 
properties, and collective phenomena can be made after the appropriate 
field redefinitions.  Specified quantum effects can be tuned for.  

A computational (e.g. computer program) can be made that 
tunes phenomena to specified materials.  For example, the band structure, 
optical properties, electron scattering, criticality involving temperature, 
and collective phenomena can be inputs and the lattice structure and external 
fields could be outputs.  This is direct to implement using the full 
effective action, if accurate from the bare Lagrangian and its interactions. 
An arbitrary number of interactions can be used in the bare theory, without 
much complication from a computational perspective including quantum effects.

In principle, desired phenomena can be obtained from the lattice and 
the external fields (involving derivatives); in practice there are technical 
obstacles. It is markable that the desired phenomena can be turned into 
the microscopic data, at bandwidth involving phase transitions at room 
temperature once the interaction $f_{s_{ij};\Lambda,g_i,T}$ functions are 
tunable, which is 
possible with the effective action.   Amorphous materials with the 
appropriate bare lattice and core interactions are also available. 

\vskip .2in 
\noindent {\it Note Added:} A computer program is being developed 
to compute large orders in the scattering and effective action.  This 
program should have obvious importance to both theory and industry.

\vfill\break

\end{document}